\documentstyle[12pt,apjpt4]{article}

\begin{document}

\title{\Huge\bf Quasi-Periodic Occultation by a Precessing Accretion Disk 
and Other Variabilities of SMC X-1}

\author{Patrick Wojdowski\altaffilmark{1}, George W.
Clark\altaffilmark{1}, and Alan M. Levine}
\vspace{0.1in}
\affil{Center for Space Research,Massachusetts Institute of
        Technology, Cambridge, MA 02139}

\author{Jonathan W. Woo}
\vspace{0.1in}
\affil{Harvard-Smithsonian Center for Astrophysics, Cambridge, MA 02138}

\author{Shuang Nan Zhang}
\vspace{0.1in}
\affil{Universities Space Research Association/Marshall Space Flight
Center, ES 84, Huntsville, AL 35812}

\altaffiltext{1}{Department of Physics, Massachusetts Institute of
        Technology, Cambridge, MA 02139}

\authoremail{pswoj@space.mit.edu}

\begin{abstract}

We have investigated the variability of the binary X-ray pulsar,
SMC~X-1, in data from several X-ray observatories.  We confirm the
$\sim 60$-day cyclic variation of the X-ray flux in the long-term
monitoring data from the {\it RXTE\/} and {\it CGRO\/} observatories.
X-ray light curves and spectra from the {\it ROSAT\/}, {\it Ginga\/},
and {\it ASCA\/} observatories show that the uneclipsed flux varies by
as much as a factor of twenty between a high-flux state when 0.71
second pulses are present and a low-flux state when pulses are absent.
In contrast, during eclipses when the X-rays consist of radiation
scattered from circumsource matter, the fluxes and spectra in the high
and low states are approximately the same.  These observations prove
that the low state of SMC~X-1 is not caused by a reduction in the
intrinsic luminosity of the source, or a spectral redistribution
thereof, but rather by a quasi-periodic blockage of the line of sight,
most likely by a precessing tilted accretion disk.  In each of two
observations in the midst of low states a brief increase in the X-ray
flux and reappearance of 0.71 second pulses occurred near orbital
phase 0.2. A similar brief flux increase near orbital phase 0.2 was
observed during a low-state observation that did not have sufficient
time resolution to detect 0.71 second pulses. These brief increases
result from an opening of the line of sight to the pulsar that may be
caused by wobble in the precessing accretion disk. The records of spin
up of the neutron star and decay of the binary orbit are extended
during 1991-1996 by pulse-timing analysis of {\it ROSAT\/}, {\it
ASCA\/}, and {\it RXTE\/} PCA data.  The pulse profiles in various
energy ranges from 0.1 to $>21$ keV are well represented as a combination
of a pencil beam and a fan beam.  Finally, there is a marked
difference between the power spectra of random fluctuations in the
high-state data from the {\it RXTE\/} PCA below and above 3.4 keV.  In
the higher energy range the spectrum has a sharp break at 3.3 Hz, with
fitted power law indices of 0.45 and 1.76 below and above the break.
No break is evident in the power spectrum below 3.4 keV, and the
fitted power law index is 0.51. In both spectra there is a positive
deviation from the fitted power law around 0.06 Hz that may be QPO.
\end{abstract}


\section{INTRODUCTION}
SMC~X-1, located in the Small Magellanic Cloud (SMC), is the most
luminous and rapidly spinning of the persistent accretion-powered
X-ray pulsars. Its X-ray flux pulses with a 0.71 second period, it is
eclipsed by its B0 supergiant companion for $\sim 15$ hours of the
3.892-day orbit, and its uneclipsed X-ray flux exhibits aperiodic
variations on time scales from tens of milliseconds to months.  Its
rotation rate is increasing on a time scale of 10${^3}$ yr, and its
orbit is decaying on a time scale of 10${^6}$ yr. With a
well-determined distance, small amount of intervening matter, and
extreme values of luminosity and spin rate, SMC X-1 is a well-placed
and particularly interesting object for detailed studies.

X-rays from ``an extended region or set of sources'' in the SMC were
discovered in a rocket observation by Price et al.~(1971).  Leong et
al.~(1971) identified a discrete source in the ``wing'' of the SMC in
{\it Uhuru\/} observations and named it SMC X-1.  Further analysis of
these and additional {\it Uhuru\/} observations by Schreier et
al.~(1972) revealed the eclipse phenomena which established its binary
nature.  The identification of its optical counterpart as the B0I star
Sk 160, suggested by Webster et al.~(1972), was confirmed by
Liller~(1973) who observed an optical intensity variation with the
same period as that of the X-ray eclipses. Lucke~(1976) discovered the
X-ray pulsations in rocket and Apollo satellite observations.
Accurate values of the orbital elements of SMC X-1 and rates of
spin-up due to accretion torque were derived by pulse-timing analysis
of {\it SAS~3\/} data by Primini, Rappaport, \& Joss~(1977) and of
{\it Ariel~V\/} data by Davison~(1977).  Measurements of the velocity
curve of Sk 160 by optical spectroscopy (\cite{hut77}; \cite{rey93}),
combined with the velocity curve of the pulsar, yielded narrow limits
on the masses of both components of the binary system.  An X-ray burst
lasting $\sim80$ s and aperiodic variability on time scales ranging
from tens of milliseconds to hours were reported from {\it EXOSAT\/}
observations by Angelini, Stella, \& White~(1991).  Levine et
al.~(1993) (hereinafter LRD) determined the rate of orbital decay in a
comprehensive analysis of the eclipse center times derived by
pulse-timing analysis of data from various satellite observations
through 1989, including their own observations with the {\it Ginga\/}
satellite.  Their publication includes a discussion of the
implications of the observed decay for the orbital dynamics of the
system.

The circumstellar environment of the Sk 160/SMC X-1 system has been
the subject of several investigations.  Van Paradijs \&
Zuiderwijk~(1977) attributed anomalies in five-color photometry of Sk
160 to the optical emission from an accretion disk around the neutron
star, and Tjemkes, Zuiderwijk, \& van Paradijs~(1986) interpreted the
optical light curve in terms of ellipsoidal variations, disk emission,
and X-ray heating effects.  Hammerschlag-Hensberge, Kallman, \&
Howarth~(1984) found evidence in UV spectra obtained from {\it IUE\/}
of the influence of X-ray illumination on the stellar wind of Sk 160
in the form of a deviation from spherical symmetry of the outflow and
an anomalously low terminal velocity compared to stars of similar
spectral type.  Blondin \& Woo~(1995) modeled the wind dynamics of the
Sk 160/SMC X-1 system in three-dimensional hydrodynamical calculations
which took account of the effects of X-ray ionization.  Model light
curves and spectra derived by Monte Carlo calculation of X-ray
propagation through the stellar wind conformed well to the light
curves and spectra of SMC X-1 derived from {\it Ginga\/} and {\it
ROSAT\/} observations, and in particular with the data obtained during
eclipse transitions when the X-ray line of sight passes close to the
surface of Sk 160~(Woo et al.~1995).
 
Extended periods of very low X-ray flux of the uneclipsed source have
been reported in observations with {\it Uhuru\/} (\cite{sch72}), {\it
Copernicus\/} (\cite{tuo75}), {\it Ariel~V\/} (\cite{coo78}), and {\it
COS-B\/} (\cite{bon81}). An extended period of exceptionally high
flux, beginning in 1970 September and lasting $\sim 100$ days, was
reported from a three-year record of observations by the {\it Vela 5B}
monitor (\cite{whi94}). Gruber \& Rothschild~(1984), in particular,
reported a large amplitude high-low state variation in the X-ray flux
in data from three $\sim 80$ day observations with the UCSD/MIT
instrument on {\it HEAO~1\/} in 1977 and 1978. They noted that the
observed variations could be ascribed to either a quasi-periodic cycle
with a duration of about sixty days or to a band-limited red noise
process of unspecified origin; a strictly periodic variation was not
consistent with their observations.  They also suggested two possible
physical mechanisms for a cyclic variation --- intrinsic luminosity
variation due to unstable mass transfer or periodic occultation by a
precessing tilted accretion disk.

In this paper we address several topics concerning the variability of
SMC X-1 and the evolution of the Sk 160/SMC~X-1 system. The
observations and data reduction are described in \S2. In \S 3, we
present long-term monitoring data from the All-Sky Monitor (ASM) on
the {\it Rossi X-ray Timing Explorer\/} ({\it RXTE\/}) and the Burst
and Transient Source Experiment (BATSE) on the {\it Compton Gamma-ray
Observatory\/} ({\it CGRO\/}) that confirm the $\sim60$-day cyclic
variability of the X-ray flux (brief reports of this result were made
by Zhang et al.~(1996) and Levine et al.~(1996)).  We then present an
analysis of the X-ray light curves and spectra derived from various
pointed-mode observations by the {\it ASCA\/}, {\it ROSAT\/}, {\it
Ginga\/}, and {\it RXTE\/} observatories for the purpose of
determining the cause of the cyclic variability. In \S 4, the records
of spin-up of the neutron star and decay of the binary orbit are
extended from mid 1991 to early 1996 with results derived by
pulse-timing analysis of data from {\it ROSAT\/}, {\it ASCA\/} and
{\it RXTE\/}. Pulse profiles in several energy ranges from 0.1 to
$>21$ keV are derived from these data, and modeled by a pulsar
radiation pattern consisting of a pencil beam and a fan beam. Finally,
in \S5 we compare the power density spectra of the random fluctuations
of the high-state flux in the high and low-energy portions of the data
recorded by {\it RXTE}.

\section{OBSERVATIONS AND INITIAL DATA REDUCTION}

Bright X-ray sources have been monitored with the BATSE instrument in
the 20-100 keV range since April of 1991 and with the ASM instrument
in the 1.5-12 keV range since 20 February 1996. The BATSE instrument
has been described by Fishman et al.~(1994), and the ASM instrument by
Levine et al.~(1996).  Source intensities are derived from BATSE data
by fitting the predicted changes in count rates caused by earth
occultations to the measured rates (\cite{har92}).  Intensities of
known sources are derived from ASM data by fitting predicted shadow
patterns cast by coded masks to observed patterns recorded by
position-sensitive proportional counters.

The {\it Ginga\/} data were obtained with the Large Area Counter (LAC)
comprised of 8 proportional gas counters with a peak total effective
area of 4000 cm$^{2}$, a mechanically-collimated field of view of
1.1\arcdeg $\times$ 2.0\arcdeg ~FWHM, an energy range from 1.5 to 36
keV, and an energy resolution of 20\% FWHM at 5.9 keV. The {\it
Ginga\/} observatory has been described by Makino~(1987) and details
of the LAC by Turner et al.~(1989).

Data from four {\it ROSAT\/} imaging observations were obtained with
the Position Sensitive Proportional Counter (PSPC) in the focal plane
of the X-ray telescope (XRT). The {\it ROSAT\/} mission has been
described by Tr\"{u}mper~(1983) and the XRT-PSPC system by Pfeffermann
et al.~(1987) and Aschenbach~(1988). The XRT-PSPC had an angular
resolution of $\sim25\arcsec$, and was sensitive in the energy range
from 0.1 to 2.5 keV.  We denote the four {\it ROSAT\/} observations
with the labels ``1'', ``2'', ``3'', and ``C''.  In the numbered
observations, SMC~X-1 was close to the center of the imaged field of
view.  Observation ``C'' was part of a survey of the SMC
(\cite{kah96}); in that observation SMC~X-1 was $\sim$40\arcmin
~off axis.

The {\it ASCA\/} imaging data were obtained with the Solid-state
Imaging Spectrometer (SIS) and the Gas Imaging Spectrometer (GIS)
which are sensitive in the energy range from approximately 0.3 to 12
keV.  The {\it ASCA\/} observatory has been described by Tanaka,
Inoue, \& Holt~(1994).

The pointed-mode {\it RXTE\/} data were obtained during the in-orbit
checkout (IOC) period of operation with the Proportional
Counter Array (PCA) which has a peak effective area of 7000 cm$^{2}$,
a mechanically-collimated 1\arcdeg ~FWHM field of view, 18\% energy
resolution at 6 keV, and an energy range from 2 to 60 keV (\cite{jah96}).

Table \ref{obs} summarizes the times and exposures of the pointed
observations and the ``state'' of SMC~X-1 in the high-low cycle as
defined below.

\subsection{Background Subtraction of Pointed Observations}

Background rates in the source region of the images produced by the
{\it ROSAT\/} and {\it ASCA\/} observatories were estimated from the
count rates in separate nearby regions. For {\it ROSAT\/} observations
1, 2, and 3 we chose the source region to be a circle of radius
150\arcsec ~and the background region to be an annulus of inner radius
150\arcsec ~and outer radius 300\arcsec.  For the poorly focused {\it
ROSAT\/} C observation, we chose the source region to be a circle of
radius 262.5\arcsec ~and the background region to be an annulus with
radii of 262.5\arcsec ~and 487.5\arcsec.  We excluded from both the
background and source regions a circular region of radius 150\arcsec
(188\arcsec ~for the off-axis pointing) centered on the foreground
coronal star HD~8191 - a weak soft X-ray source.  Also, we excluded a
circular region of radius 150\arcsec ~around the transient source RX
J0117.6-7330 (Clark et al. 1997) in the two observations (2 \& 3) in
which it appeared.

In the GIS image of the {\it ASCA\/} observations, we chose the source
region to be a circle of radius 383\arcsec ~and the background region
to be an annulus of radii 383\arcsec ~and 766\arcsec.  In observation
2, a circular region of radius 150\arcsec ~around HD~8191was removed
from both the source and background regions.  In the SIS image, we
chose the source region to be a square 380\arcsec ~on a side centered
on SMC~X-1 and used the remainder of the chip as the background
region.  HD~8191 was not in the field of view of the SIS in these
observations.

Background corrections to the {\it Ginga\/} data were made according
to the algorithm of Hayashida et al.~(1989).  Background corrections
to the {\it RXTE\/} data were derived from the earth-occulted count
rates which are listed in Table \ref{pca_rates} together with the
on-target rates in seven energy ranges. We note that the {\it ROSAT\/}
images show several faint and soft sources near SMC~X-1 that were
within the {\it Ginga\/} and {\it RXTE\/} fields of view.  No
corrections to the background rates were made for these sources
because their total contribution to the {\it Ginga\/} count rate was
less than a few percent of the SMC~X-1 eclipsed count rate
(\cite{woo95a}), and less than one percent of the {\it RXTE\/} rates.

\section{QUASI-PERIODIC HIGH-LOW FLUX VARIATION}
\subsection{Long-Term Monitoring Observations}
Figure 1 displays the X-ray light curves of SMC~X-1 measured by the
ASM and BATSE instruments during times when the pulsar was uneclipsed
and in the orbital phase range between 0.1 and 0.9 relative to
eclipse center. The ASM data are binned in intervals of the 3.89-day
orbital period, and the BATSE data in five-day intervals.

A cyclic flux variation on a time scale of $\sim 60$ days is clearly
evident in the ASM light curve.  To quantify the duration of each
cycle, which appears to be variable, we defined times of the beginning
and end of each high state to be when the count rate
increased/decreased to 2 counts s$^{-1}$.  We determined these times
by interpolating between successive points with values below and
above, or above and below, 2 counts s$^{-1}$.  The times between
successive rises and falls defined in this way are plotted against
cycle number in Figure 2 where they are seen to range from below 50 to
above 60 days with a generally decreasing trend. The ASM light curve
shows that the flux often increased relatively quickly at the
beginning of each high state and decreased more slowly at the end.

To test for the presence of cyclic variability in the BATSE data, we
divided the data into four subsets of equal duration and folded each
subset into 5 phase bins with trial periods between 30 and 100 days.
In this process, the mean flux and its standard deviation in each
phase bin were calculated using the inverses of the variances of the
individual measurements as weights.  At each trial period, we
calculated the reduced $\chi^2$ statistic of the light curve assuming
that no signal was present.  The values of the reduced $\chi^2$
statistic are plotted in Figure 3 for each quarter of the BATSE light
curve and for the entire BATSE light curve. Also shown are the results
of the same analysis applied to the entire ASM light curve.  Leahy et
al.~(1983) showed that if a sinusoidal signal with a fixed period,
$P$, is present in the light curve, such a folding analysis will show
a peak centered at that period with a FWHM of $P^2/T$, where $T$ is the
length of the observation.  In Figure 3 the horizontal bars show the
expected FWHM of a sinusoidal signal with a 55-day period.  We note
that the positions of the peaks shift and some of peaks are
substantially wider than the horizontal bars. These features, together
with the drift of the time since last increase/decrease in the ASM
light curve (Figure 2), are strong evidence for the absence of a
long-term coherent periodicity in these data sets.  Instead, the
observed cyclic variation can be best characterized as quasi-periodic
with recurrence times that wander between $\sim$50 and $\sim$60 days.
Attempts to examine the variation of the cycle more closely in the
BATSE light curve by smoothing and by correlation with a sine wave
were inconclusive, presumably due to the low signal-to-noise ratio in
the data. 

\subsection{Short-Term Pointed Observations}
\label{pointed_obs}
Background-subtracted count rates, derived from the {\it Ginga\/}, {\it
ROSAT\/}, {\it ASCA\/}, and {\it RXTE\/} PCA observations, are plotted in
Figure 4 against orbital phase.  We multiplied the
background-subtracted count rates of the off-axis {\it ROSAT\/}
observation by a factor of 1.7 to compensate for the reduction in
effective area relative to the on-axis observation.  So that the {\it
Ginga\/} and {\it ASCA\/} count rates may be directly compared, we show
the count rates from the pulse height channels which correspond to the
energy range 1-10 keV.  For the {\it ASCA\/} GIS, these are PI channels
85 through 850.  For the {\it Ginga\/} LAC these are channels 3 through
18 in the MPC-2 mode data and 1 to 4 in the MPC-3 mode data. 

With the exception of the {\it Ginga\/} observation, which occurred
before the beginning of the BATSE light curve, all of the pointed
observations discussed in this paper occurred during the time of the
BATSE observations, and before the {\it RXTE\/} ASM observations.
Unfortunately, the signal-to-noise ratio of the BATSE light curve is
not large enough to allow certain determination of the phases in the
flux cycle at which the individual pointed observations occurred.
Furthermore, because of the quasi-periodic nature of the cycle, it is
not possible to extrapolate backwards from the ASM light curve.
However, the ASM light curve shows that the X-ray flux averaged over
the uneclipsed portion of each binary orbit rises and falls smoothly
over each cycle.  Therefore, we are confident that the observed
out-of-eclipse flux of the source, as it is revealed in the light
curves, is a reliable indicator of the state of the source.

The contrast between the high and low states of SMC~X-1 is most
clearly demonstrated by the various {\it ROSAT\/} observations.  Where
observation 1 and the region C observation overlap in orbital phase
outside of eclipse, they differ in (corrected) count rate by a factor
of at least 20.  We therefore identify {\it ROSAT\/} observation 1 as
having occurred during the high state and the region C observation as
having occurred during the low state.  The light curve of {\it
ROSAT\/} observation 2 behaves like that of the region C observation
where the two overlap so we identify this observation as having
occurred during the low state as well.  Outside of eclipse, {\it
ROSAT\/} observation 3 shows a count rate comparable to that of
observation 1, so we identify it as having occurred during the high
state.

The two {\it ASCA\/} observations also differ by a factor of about
twenty in their out-of-eclipse count rates, so we assume {\it ASCA\/}
observation 1 occurred during the high state and {\it ASCA\/}
observation 2 occurred during the low state.  However, the decline in
the count rate in observation 1 beginning near orbital phase 0.5 may
be a transition to the low state.  In the {\it Ginga\/} observation,
the out-of-eclipse behavior and count rate is comparable to {\it
ROSAT\/} observation 1 and {\it ASCA\/} observation 1, so we conclude
it also occurred during the high state. The brief {\it RXTE\/} PCA
observation near orbit phase 0.5 clearly occurred during a high state.

Interesting features of the low-state light curves derived from the
{\it ROSAT\/} observations 2 and C are the increases in the count
rates near orbital phase 0.25.  In both cases, 0.71 second pulsations
are present during the increase.  A similar turn-on may be seen in the
light curve (not shown herein) derived from the {\it ROSAT\/} All-Sky
Survey (\cite{kah96}), although those data do not have sufficient time
resolution to permit a test for the presence of pulsations.  {\it
ROSAT\/} observation 2 ends near phase 0.3 with this increase.  The
{\it ROSAT\/} C observation has a gap from phase 0.3, where the count
rate is increasing, to phase 0.58 (plotted at phase -0.42), where the count
rate is lower than before.

In Figure 5 we plot the photon number spectra of SMC~X-1 from all the
observations during the in-eclipse portions of the binary orbit (orbital
phase -0.07 to +0.07).  For the purpose of comparing these spectra, we
fitted all of them simultaneously with the following model spectrum that
consists of the sum of a black-body spectrum and a power law with a
high energy cutoff together with interstellar absorption:
\begin{equation}
I(E) = \exp[-\sigma(E)N_{\rm H}] [f_{\rm bb}(E)+f_{\rm pl}(E)],
\label{eq:s1spfunc}
\end{equation}
where
\[
\begin{array}{ll}
f_{\rm bb}(E) & = I_{\rm bb}(E/E_{\rm bb})^{2}(e-1)[\exp (E/E_{\rm bb})-1]^{-1}, \\
f_{\rm pl}(E) & = I_{\rm pl} (E/\rm 1 keV)^{-\alpha}\times\left\{ 
   \begin{array}{ll} 
        1,                              &  E < E_{\rm c} \\
        \exp{[-\frac{E-E_{\rm c}}{E_{\rm f}}]}, &  E\geq E_{\rm c}
    \end{array} \right\}. \\
\end{array}
\]
$I_{\rm bb}$ is the the intensity of the blackbody component at
$E=E_{\rm bb}$, $I_{\rm pl}$ is the intensity of the power law
component at $E=1$~keV, and $\sigma(E)$ is the photoelectric
absorption cross section of cold matter (\cite{mor83}).  The results
of the fit are given in Table \ref{pars}.  The convolutions of this
model with each of the instrument responses are plotted in Figure 5
along with the corresponding observed spectra.  The instrument
response for the {\it ROSAT\/} C observation includes the reduced
efficiency at the off-axis location.

The high-low state identifications we have inferred from the observed
fluxes leave no doubt that the spectra displayed in Figure 5 include
instances from both the high and low states during the eclipse when
the only X-rays detected are those that have been scattered from
circumsource matter.  Since these five spectra show no more than a
factor of two difference in flux at any X-ray energy, we conclude that
the average illumination of the circumsource matter varied little
between the high and low states.  Moreover, all five spectra have
similar shapes.  Thus, the cause of the cyclic variation cannot be
either a variation in the intrinsic luminosity of SMC~X-1 or a
redistribution of the luminosity across the spectrum. We conclude that
the cause of the cyclic variation between states of high and low flux
is a quasi-periodic blockage of the line of sight. The same conclusion
regarding the cause of the long-term cyclic variation in the X-ray
flux of LMC~X-4 has been drawn on the basis of similar evidence
(\cite{woo95b}). 

\subsection{Motions of the Accretion Disk}

Shortly after discovery of the 35-day high-low state variation in Her
X-1, Katz~(1973) suggested that if the rim of the neutron star's
accretion disc does not lie in the plane of the binary orbit, then the
plane of the rim will precess due to the gravitational force of the
companion star. Katz argued that if the rim is sufficiently tilted, it
will periodically block the line of sight to the neutron star, causing
the observed cyclic high-low variation.  His expression for the forced
precession period, $P_{\rm f}$, of a ring of non-interacting particles
can be written in the form
\begin{equation}
P_{\rm f}=\frac{4}{3}P_{\rm b}\left(\frac{R_{\rm L}}{R_{\rm r}}\right)^{3/2}
\left[\frac{0.6q^{2/3}+\ln (1+q^{1/3})}{0.49q^{2/3}}\right]^{3/2}
(q+q^{2})^{1/2}(\cos\beta)^{-1},
\label{Pf}
\end{equation}
where $P_{\rm b}$ is the binary orbital period, $R_{\rm L}$ is the
volume-averaged radius of the Roche lobe, $R_{\rm r}$ is the ring radius, $q$
is the ratio of the neutron star mass to the companion star mass, and
$\beta$ is the tilt angle of the ring with respect to the plane of the
binary orbit. The quantity in square brackets is an approximate
expression for the ratio of the binary separation to the Roche lobe radius
(Eggelton 1983).  

Analyses of the optical light curves of Sk~160/SMC~X-1 indicate that
the accretion disk radius is $\sim0.7$ to $1.0$ times the Roche lobe
radius (\cite{how82}; \cite{tje86}; \cite{khr87}). Setting $R_{\rm
r}/R_{\rm L}=0.7$, $q=0.0909$ (\cite{rey93}), $P_{\rm b}=3.892$ days,
and $\cos\beta\approx 1$, we find the precession period predicted by
equation \ref{Pf} is 31 days or 0.56 times the typical observed cycle
time of 55 days. The optical light curves of the two other X-ray
binaries with clear quasi-periodic high-low state cycles, Her~X-1 and
LMC~X-4, also indicate that their accretion disks nearly fill the
Roche lobes of their neutron stars (\cite{ger76}; \cite{ilo84};
\cite{hee89}).  The corresponding precession periods (assuming $R_{\rm
r}/R_{\rm L}=0.7$ and $\cos \beta\approx 1$) are 0.56 and 0.35
times the observed high-low cycle periods of Her~X-1 and LMC~X-4,
respectively. Thus, in all three cases a simple model of an occulting
rim of non-interacting particles predicts precession periods
substantially less than the observed high-low cycle times.

Numerous theoretical investigations of the effects of the accretion
rate, viscosity, and X-ray illumination on the creation of warp and
tilt in the disk and on the precession rate have been carried out.
Among these investigations we note in particular the work of Iping \&
Petterson~(1990).  They made numerical simulations with results that
showed stable precessing disks with quasi-periods that could be
adjusted to fit observations by appropriate choices of the accretion
rate, the conversion efficiency with which accretion energy is
converted into X-rays, and the viscosity parameters. Pringle~(1996)
confirmed their results with a more rigorous treatment, and showed
that a warp instability due to radiation pressure will develop outside
a critical radius which is $\approx3\times10^8$~cm for a neutron star.
The radius of the Roche lobe in SMC~X-1 is $\approx 2.9\times
10^{11}$cm so the accretion disk is clearly large enough to be
unstable to warping. Thus it appears that a detailed model of a
precessing tilted accretion disk could provide a quantitative
explanation for the observed high-low cycle of SMC~X-1.

It is unlikely that random flares were the cause of the short-term
flux increases observed near orbital phase 0.25 in the three
low-states observations.  The presence of 0.71 second pulses in the
{\it ROSAT\/} 2 and C observations shows that the increased flux came
directly from the neutron star. It is also unlikely that the increase
in the {\it ROSAT\/} C observation was the beginning of a high-state
because a decrease followed soon after, and because the observation
occurred only ten days after the source was observed to be in a high
state in {\it ROSAT\/} observation 1.  The ASM light curve shows that
low states typically last for 15-20 days.  Thus the {\it ROSAT\/} C
observation probably occurred in the early part of a low state.  These
increases are reminiscent of the low-state turn-ons in Her~X-1 which
occur near orbital phases 0.2 and 0.7 (cf. Fig. 2 of Priedhorsky \&
Holt 1987), and which have been attributed (\cite{lev82}) to a wobble
which may be a common feature of precessing accretion disks.  Further
observations of SMC~X-1 are  needed to determine the regularity and
nature of its low-state turn-ons.

\section{PULSE-TIMING ANALYSIS OF SPIN UP AND ORBITAL DECAY}
\subsection{Observations and Analysis}

We carried out pulse timing analysis of the {\it ROSAT\/}, {\it ASCA\/},
and {\it RXTE\/} PCA data to extend previously published results on the
spin up of the neutron star and decay of the binary orbit through
1996 January. The {\it ROSAT\/} and {\it ASCA\/} data consist of the energy,
image position, and readout time of each photon detected in an image
of SMC~X-1. The time resolution of the {\it ROSAT\/} data is a few ms,
and the time resolution of the {\it ASCA\/} data is 1/16 s.  The {\it
RXTE\/} data consist of the counts accumulated by the PCA in rebinned
time intervals of 1/64 s and five energy channels from 2 to $>21$ keV.

None of our six pointed observations span a sufficient portion of the
SMC~X-1 orbit to permit an independent determination of all of the
orbital parameters and the rate of change of the pulse frequency. The
{\it ROSAT\/} 2 and C observations had such brief exposures to the
pulsed flux that the data were only sufficient to constrain the pulse
frequency when the orbital phase (measured from eclipse center for a
circular orbit) was fixed at the value inferred from the other
observations. We restricted the timing analysis of the other four
observations to determinations of the intrinsic pulse frequency and
orbital phase at the start of each observation, while the other
orbital parameters were fixed at the values derived by LRD from
observations of longer duration. From the orbital phases we derived
the center times of the preceding eclipses.  These were combined with
the previous eclipse center times quoted by LRD in a redetermination
of the orbital decay rate.

To obtain initial estimates of the orbital phase, $\phi_{0}$, and
intrinsic pulse frequency, $f$, at the start of an observation, we
searched a two-dimensional array of phase and frequency for the pair
of values that maximized the fractional amplitude of the folded pulse
light curve derived from the entire set of solar system barycentric X-ray
arrival times corrected for orbital delay. The delay-corrected arrival
time of the $i$th photon, $t_{i}'$, was derived from its observed
arrival time, $t_{i}$, by iterative solution of the equation
\begin{equation}
        t_{i}' = t_{i}-a_{\rm x}\sin i\cos (2\pi\frac{t_{i}'}{a_{1}}+\phi _{\rm 0})
\end{equation}
where the quantities $a_{\rm x}\sin i$ and $a_{1}$ were fixed at
the values found by LRD and listed in Table \ref{timing_results}. The
change in these values, implied by the rate of orbital decay measured
by LRD, had a negligible effect on the delays computed for the present
data.

We refined the estimated values of $\phi_{0}$ and $f$ by the following
procedure aimed at obtaining a sequence of accurate pulse arrival time
delays to which calculated orbital delays could be fitted by
adjustment of the orbital phase. A portion of the delay-adjusted data
was folded at the estimated frequency, and a 12-term Fourier series
was fitted to the resulting pulse profile to form a template profile.
The data were then divided into a number of subsets, and each subset
was folded with the estimated frequency to obtain its pulse profile.
Each profile was cross-correlated with the template to determine its
relative phase shift due to errors in the estimated values of
$\phi_{0}$ and $f$.  From the phase shift of the $j$th profile we
computed the difference between the actual corrected arrival time of a
pulse near the mean arrival time, $T_{j}$, of the data subset and the
expected corrected arrival time based on the estimated frequency.  We
call $\tau(T_{j})$ the sum of that difference and the orbital delay at
$T_{j}$ calculated with the estimated $\phi_{0}$.  The quantity
$\tau(T_{j})$ is equal to the sum of the true orbital delay and the
time equivalent, $(T_{j}-T_{0})\delta f/f$, of the portion of the phase
shift due to the error, $\delta f$, of the estimated value of $f$.  A
least-squares fit of calculated values of $\tau(T_{j})$ to the set of
measured values was made by adjustment of $\phi_{0}$ and $\delta f$.
This procedure was repeated, each time using improved estimates of the
orbital phase and frequency to compute a new template and new values
of $\tau(T_{j})$ until there was no further improvement in the
least-squares fit, and no significant trend away from a zero shift in
pulse phase between the template and the subset pulse profiles.

In Figure 6 we show plots of the final pulse arrival time delays
together with the fitted curves and deviations of the fits.  The {\it
ROSAT\/} 1 and 3 and the {\it ASCA\/} 1 observations yielded
well-constrained values of the intrinsic pulse period $P=1/f$ and
eclipse center times.  In spite of the high statistical accuracy of
the {\it RXTE\/} data, the shortness of the observation resulted in
relatively large and correlated uncertainties in the eclipse center
time and intrinsic pulse frequency.

The quadratic function of orbit number defined by LRD was fitted to
the set of eclipse center times which includes the four new ones
derived in this work and the seven previous ones listed in their paper
and reproduced in Table \ref{timing_summary}.  The result is shown in
Figure 7 as a plot of the measured deviations from a linear function
of orbit number against time. The fitted values of the coefficients
are listed in Table \ref{timing_results} together with the previous
values of LRD with which they are entirely consistent.

The pulse periods from this and previous studies are plotted against
time in Figure 8 together with the deviations from a fitted quadratic
function of time with the coefficients listed in Table \ref{per_fit}.
We note that every measurement of the pulse period of SMC~X-1 has
yielded a value lower than the previous measurement.  Although the
magnitude of the rate of change of the period appears to be
decreasing, the absence of any observed episodes of torque reversal
indicates that the accretion flow has never slowed enough to allow the
inner edge of the accretion disk to retreat to the corotation radius
where braking action would be expected to occur. 

\subsection{Variation of the Pulse Profile with Time and Energy}
Figure 9 displays background-subtracted pulse profiles in five energy
ranges from $0.1$ keV to $>21$ keV derived from the {\it ROSAT\/},
{\it ASCA\/}, and {\it RXTE\/} PCA data.

Striking features of the profiles are the simplicity and smoothness of
the profiles, and the steady increase with energy in the relative
amplitude of the narrower component of the double pulse.  In marked
contrast, the profiles of various other X-ray pulsars such as Vela X-1
and LMC X-4 are complex and vary widely over the same energy range.
Such complexity may be caused by a multi-pole structure of the
magnetic field that guides the accretion flow to the surface of the
neutron star, giving rise to a multiplicity of hot spots and
complicated patterns of emission from multiple accretion columns. In
SMC~X-1 it appears that the guiding field may be represented to a good
approximation by a simple dipole inclined with respect to the spin
axis. The different energy dependence of the two components, and the
fact that the amplitude of the narrower one changes from less to more
than that of the broader one precludes the possibility that they are
caused by identical pencil beams emanating in opposite directions from
the magnetic poles.  Instead, the narrower pulse may be attributed to
the pencil beam emanating from the accretion column above one pole,
and the broader pulse to the fan beams emanating from the accretion
columns above the two poles.
 
A second prominent feature is the variability with time and energy of
the narrower pulse.  In the energy bands where data from {\it ASCA} in
April 1993 and from {\it RXTE} in January 1996 can be compared, there
is a substantial change in the relative amplitudes of the two pulses
which appears to be due to a change in the flux in the narrower pulse.
Similar variability of the narrower pulse was reported by LRD in {\it
Ginga} data from 1987, 1988, and 1989. In each of the three data sets
displayed in Figure 9 the amplitude of the narrower pulse increases
with energy, starting near zero in the {\it ROSAT} data below 1 keV,
and rising above that of the wider pulse in the {\it RXTE} data above
21 keV.

X-ray production and radiative transfer in magnetized plasma has been
computed by Nagel~(1981) for slab and cylindrical geometries.
In one particular case of cylindrical geometry with the magnetic field
parallel to the axis, which serves as a model of an accretion
column, the resulting radiation pattern has the form of combined
pencil and fan beams, i.e., the escaping ordinary photons form a pencil
beam with a maximum near 0\arcdeg ~with respect to the axis, and the
extraordinary photons form a fan beam with a maximum at 90\arcdeg.
It may be plausible, therefore to model the SMC~X-1 pattern as such a
combination. If the broad pulse is identified with a fan
beam, then its flat top can be attributed to the prolonged dwell near
the minimum angle between the line of sight and the plane
perpendicular to the dipole axis. If the narrow pulse is attributed to
the pencil beam, its energy dependence and variability may be due to
the energy dependence and variability of the opacity of the magnetized
plasma through which the pencil beam propagates.

In Figure 10 we display the observed pulse profile of
background-subtracted data from 2 to 21 keV (histogram) together with
a model profile (crosses) computed for a radiation pattern consisting
of an isotropic component plus combined pencil and fan beams. The
pencil beam flux is represented by a Gaussian function of the angle,
$\rho$, between the line of sight and the dipole axis, the fan beam
flux by a Gaussian function of $90\arcdeg-\rho$.  We call $\Theta$ the
inclination of the spin axis to the line of sight, and $\Psi$ the
angle between the spin axis and magnetic dipole axis.  Then
\begin{equation}
\rho=\cos^{-1}[\sin \Theta \cos (\phi-\phi_{0}) \sin \Psi +\cos \Theta\cos \Psi].
\end{equation}
where $\phi$ is the pulse phase. The predicted flux was computed by
the equation
\begin{equation}
F(\phi)=F_{\rm p}\exp\left(-\frac{\rho^{2}}{2w_{\rm p}^{2}}\right)+F_{\rm
f}\exp\left(-\frac{(90\arcdeg-\rho)^{2}}{2w_{\rm f}^{2}}\right)+F_{0}.
\end{equation}
We set $\Theta$ to $60^{\circ}$, which implies equality between the
the inclination of the spin axis and estimates of the orbit
inclination, and $\Psi$ to $27 ^{\circ}$.  This combination of angles
achieves the condition of a dwell of the fan beam near the line of
sight. The other parameters were adjusted by eye and are listed in
Table \ref{per_fit}.  We note that the pencil beam in this model is
similar in angular width to that in the theoretical model of
Nagel~(1981) mentioned above. On the other hand, the fitted fan beam
is substantially narrower than the computed one.

\section{RANDOM INTENSITY FLUCTUATIONS}
Power spectra of the random fluctuations in the {\it RXTE\/} PCA data
in two energy ranges are displayed in the right-hand panels of Figure
11.  The spectra were computed as follows: the data were divided into
53 segments of 256 seconds duration, the power spectrum of each
segment was computed with the normalization of Leahy et
al.~(1983), the Poisson noise component of 2 was subtracted, and the
result was divided by the mean count rate of the segment to obtain
fractional RMS normalized powers. The 53 spectra were then averaged,
and the powers due to five harmonics of the pulsed power, computed
according to the method of Angelini et al.~(1991), were subtracted.
Finally, the resulting power spectra were logarithmically rebinned for
display. The broken power laws and harmonic components that were
fitted to the normalized spectra are displayed in the left-hand panels
of Figure 11.  The errors of the normalized powers before rebinning
were all set equal to the expected RMS Poisson fluctuations computed
as the quantity
\begin{equation}
        \frac{2}{M}\left[\displaystyle \sum_{i=1}^{M}\left(1/c_{i}^{2}\right)\right]^{1/2},
\end{equation}
where $c_{i}$ is the count rate during the $i$th data segment, and $M$
is the number of segments.
        
A marked difference is apparent between the forms of the power
spectrum in the low and high energy channels. The spectrum in the
2.0-3.4 keV range is fitted by a single power law with an index of
0.51; the 3.4-21 keV spectrum conforms to a broken power law with
indices of 0.45 and 1.76 with a sharp break at 3.3 Hz.  Both spectra
show a positive deviation from the power-law fit around 0.06 Hz which
may be attributed to quasi-periodic oscillation (QPO).

The break we find in the power spectrum above 3.4 keV is similar to
those found by Takeshima ~(1994) in {\it Ginga} observations of
several pulsars. He fitted the power spectra with broken power laws,
and found breaks occurring near the fundamental pulsar frequencies.
In his analysis of data from the second and third {\it Ginga\/}
observations of SMC~X-1 listed in Table \ref{timing_summary}, he found
breaks at 2.93 Hz and 1.79 Hz.

The difference between the power spectra that we find in the high and
low-energy portions of the X-ray flux may be a clue to the effect of
magnetic field intensity on the frequency spectrum of emission
fluctuations. The flux in the low-energy channel may be dominated by
radiation emitted from the accretion disk where the magnetic field is
relatively weak.  Most of the flux in the high-energy channels
certainly arises from the accretion column and hot spot near the
surface of the neutron star, where the magnetic field is strong. The
strong field may suppress the fluctuations responsible for
low-frequency red noise, thereby causing the abrupt decrease in the
logarithmic slope of the power spectrum below 3.3 Hz.

\section{Summary}
The principal results in this paper are as follows:

The ASM light curve of SMC~X-1 confirms the presence of a high-low
state cyclic variation of the X-ray flux, and shows that the cycle
duration varied from $\sim 50$ to $\sim 60$ days. The BATSE data show
that the cyclic variation is a persistent phenomenon.

Results from pointed observations show that the uneclipsed flux of
SMC~X-1 in the high state is more than an order of magnitude greater
than in the low state.  In contrast, the eclipsed flux, consists of
X-rays scattered from circumsource matter, differs by no more than a
factor of $\sim 2$ between the two states, and the spectrum shape does
not change significantly. Thus, the high-low cyclic variation is not
caused by either a change in the luminosity of the source, or a
redistribution of the energy. Rather, it must be the result of a
quasi-periodic blockage of the line of sight.

In a simple model of occultation by a ring of non-interacting
particles at the rim of a tilted accretion disk, the forced precession
caused solely by the gravity of the companion star would have a period
of about half the observed long-term cycle duration.  Similar
differences are found between the predicted precession periods for the
simple model and the observed long-term cycle periods of Her~X-1 and
LMC~X-4. It has been shown that more detailed and realistic models of
precessing accretion disks that take account of the effects of the
accretion rate, accretion energy conversion efficiency, and viscosity
in the disk, can be adjusted to cause model disks to tilt and precess
with periods in agreement with the observed long-term cycles. We
therefore conclude that the high-low state cycle in the X-ray flux
from SMC~X-1, like the similar cycles of Her~X-1 and LMC~X-4, is the
result of quasi-periodic occultation by a precessing tilted accretion disk.

Timing analysis of the 0.71 second pulses shows a continuation of the
decay of the binary orbit through 1996 January at a rate consistent
with previous results.

The pulse period has continued its apparently monotonic decrease
at a gradually decreasing rate.

The pulse profile can be modeled with fair accuracy by a three
component radiation pattern consisting of an isotropic component plus
combined pencil and fan beams symmetric about an axis inclined at
27\arcdeg~from the spin axis, with the spin axis inclined at
60\arcdeg~to the line of sight.

The power spectrum of the flux recorded by {\it RXTE} in energy
channels below 3.4 keV is fitted by a single power law with index 0.51
over the frequency range from 0.01 to 32 Hz.  In marked contrast, the
power spectrum in energy channels above 3.4 keV is fitted by a broken
power law with a sharp decrease in index from 1.76 above 3.3 Hz to
0.45 below. The low-energy flux may be dominated by radiation emitted
from the accretion disk where the magnetic field is relatively weak.
The high-energy flux is emitted primarily in regions of strong
magnetic field which may suppress the emission fluctuations that cause
the low-frequency red noise.

In both the high and low-energy power spectra there is a positive
deviation from the fitted power law around 0.06 Hz that may be QPO.

\acknowledgments

We used the {\it XSPEC\/} X-ray spectral fitting program for our
spectral analysis.  Archival {\it ROSAT\/} and {\it ASCA\/} data was
obtained with the use of the High Energy Astrophysics Science Archive
Research Center Online Service, provided by the NASA/Goddard Space
Flight Center.  We thank Jane Turner and Ken Ebisawa of the {\it user
support teams for advice on reducing and analyzing ROSAT\/} and {\it
ASCA\/} data, and Edward Morgan of MIT for assistance in the use of
{\it RXTE} data acquired during the in-orbit calibration (IOC) period
of operation. We specially thank the referee, J. van Paradijs, for his
careful reading and valuable suggestions for improvements in the
manuscript.

\begin{deluxetable}{llrrrcr}
 \tablecaption{Summary of Pointed Observations}
\tablehead{
\colhead{Instrument and ID}   &  \colhead{Start Time-Stop Time}  &
\colhead{Phases\tablenotemark{a}} &
\colhead{Exposure (s)}  &  \colhead{State} & \colhead{ref}}
\startdata
{\it Ginga\/}  & 1989 Jul 30.78-Aug 3.20\tablenotemark{b} &  0.34 --  0.21 & 74240 & high & 1\\
{\it ROSAT\/} 1& 1991 Oct 7.17-8.13      &  0.48 -- -0.29 & 16634 & high & 2\\
{\it ROSAT\/} C& 1991 Oct 16.31-19.98    & -0.18 -- -0.24 & 22173 &  low & 3\\
{\it ROSAT\/} 2& 1992 Sep 30.71-Oct 2.59 & -0.15 --  0.33 &  8932 &  low & 2\\ 
{\it ROSAT\/} 3& 1993 Jun 2.99-4.20      & -0.13 --  0.18 & 11864 & high & 2\\ 
{\it ASCA\/} 1 & 1993 Apr 26.94-27.97    &  0.35 -- -0.39 & 34507 & high &  \\ 
{\it ASCA\/} 2 & 1995 Oct 18.79-19.88    & -0.17 --  0.11 & 30628 &  low &  \\
{\it RXTE\/}   & 1996 Jan 11.00-11.23    &  0.47 -- -0.47 & 14536 & high &  \\ 
\tablenotetext{a}{Binary orbit phase range on the (-0.5,0.5) interval.  All
  observations span less than one binary orbit.}
\tablenotetext{b}{SMC~X-1 was observed for more than one binary
  orbit.  Only a subset of the data actually collected is described here.}
\tablerefs{
(1) Woo, et al. 1995; (2) Clark, et al. 1997; (3) Kahabka \& Pietsch 1996.}
\enddata
\label{obs}
\end{deluxetable}
\clearpage

\begin{deluxetable}{ccc}
\tablewidth{3.5in}
\tablecaption{Average Count Rates in Seven Energy Ranges of {\it RXTE\/}
Data}
\tablehead{
\colhead{}              &\multicolumn{2}{c}{Average Count Rates} \\     
\colhead{}              &\multicolumn{2}{c}{(counts s$^{-1}$)}          \\      
\cline{2-3}                                                     \\
\colhead{Energy Range}  &\colhead{SMC~X-1} &\colhead{Earth}     \\
\colhead{(keV)}         &\colhead{+Background}&\colhead{Occulted}       }
\startdata
2.0-3.4         &43.5                   &5.7                    \\
3.4-5.8         &173                   &5.0                    \\
5.8-8.6         &173                    &4.6                    \\
8.6-21          &238                    &21                    \\
21-60           &74                     &53                     \\
\enddata
\label{pca_rates}
\end{deluxetable}
\clearpage

\begin{deluxetable}{ll}
\tablewidth{4.0in}
\tablecaption{Fitted Values of the Spectral Function Parameters}
\tablehead{
\colhead{Parameter}     &\colhead {Value}{(1$\sigma$ error)}    }
\startdata
$N_{\rm H}$(10$^{20}$~cm$^{-2}$)               \dotfill\ & 5.9$\;\;\;\;\;\;$(4.2) \\
$E_{\rm bb}$ (keV)                             \dotfill\ & 0.25$\;\;\;\;$(0.02)\\
$I_{\rm bb}$ ($10^{-3}$ photons cm$^{-2}$ s$^{-1}$ keV$^{-1}$)\dotfill\
                                                         & 9.$\;\;\;\;\;\;\;\;$(3)   \\
$\alpha$                                       \dotfill\ &-0.2$\;\;\;\;\;$(0.1) \\
$I_{\rm pl}$(10$^{-5}$~photons cm$^{-2}$ s$^{-1}$ keV$^{-1}$)
\dotfill\ 
                                                         & 9.8$\;\;\;\;\;\;$(3.1)  \\
$E_{\rm c}$ (keV)                              \dotfill\ & 6.5$\;\;\;\;\;\;$(0.4)  \\
$E_{\rm f}$ (keV)                              \dotfill\ & 5.9$\;\;\;\;\;\;$(0.4)  \\
$\chi_{\nu}^{2}$                               \dotfill\ & 3.96       \\
\enddata
\label{pars}
\end{deluxetable}
\clearpage

\begin{deluxetable}{lllll}
\tablecaption{Summary of Mid-Eclipse Times and Pulse Period Measurements}
\tablehead{
\colhead{Observatory}   &\colhead{$T_{\pi/2}^{a}(1\sigma)$} &\colhead{$N$} 
&\colhead{$P_{\rm pulse}(1\sigma)$}     &\colhead{Reference}    \\
\colhead{}              &\colhead{(MJED)}                       &\colhead{}
&\colhead{s}                            &\colhead{}}    
\startdata      
{\it Uhuru\/}     &40963.99(2)    &$-481$ &\dotfill       &Schreier et al. 1972  \\
{\it Uhuru\/}     &41114.2\tablenotemark{b}
&\dotfill&0.71748(26)   &Henry \& Schreier 1977 \\
Aerobee         &41999.6\tablenotemark{b}       &\dotfill&0.7164(2)     &Yentis et al. 1977     \\
{\it Copernicus\/}&42275.65(4)    &$-144$ &\dotfill       &Tuohy \& Rapley 1975   \\
{\it Apollo-Soyuz\/}&42613.8\tablenotemark{b}     &\dotfill&0.7151(2)     &Yentis et al. 1977     \\
{\it SAS 3\/}     &42836.1828(2)  &0      &0.71488585(4)  &Primini et al. 1977    \\
{\it Ariel V\/}   &42999.6567(16) &42     &0.7147337(12)  &Davison 1977           \\
{\it Einstein\/}  &43985.907\tablenotemark{b}     &\dotfill&0.713684(32)  &Darbro et al. 1981     \\
{\it COS B\/}     &43116.4448(22) &72     &\dotfill       &Bonnet-Bidaud et al. 1981 \\
{\it Ginga\/}     &46942.47237(15)&1055   &0.710592116(36)&LRD                  \\
{\it Ginga\/}     &47401.744476(7)&1173   &0.7101406720(15)&LRD                 \\
{\it Ginga\/}     &47740.35906(3) &1260   &0.709809901(32)&LRD                  \\
{\it ROSAT\/} 1   &48534.34786(35) &1464   &0.70911393(34) &Present work           \\
{\it ROSAT\/} 2   &48892.4191\tablenotemark{b} &1556 &0.7088166(6) &Present work          \\
{\it ASCA\/}      &49102.59109(82)        &1610   &0.7086390(9)   &Present work           \\
{\it ROSAT\/} 3   &49137.61911(50)                &1619   &0.7085992(5)   &Present work           \\
{\it RXTE\/}       &50091.170(63)         &1864   &0.70767331(8)  &Present work           \\
\enddata
\tablenotetext{a}{Mid-eclipse time, equivalent to time when mean
longitude $l$ equals $\pi/2$ for a circular orbit (c.f. \cite{dee81}).}
\tablenotetext{b}{Set to predicted mid-eclipse times}
\label{timing_summary}
\end{deluxetable}
\clearpage

\begin{deluxetable}{lll}
\tablewidth{5.0in}
\tablecaption{Parameters of the Spin-Up and Orbital Decay}
\tablehead{
\colhead{}              &\multicolumn{2}{c}{Value ($1\sigma$ uncertainty)}\\
\cline{2-3}                                                     \\
\colhead{Parameter}     &\colhead{Present Work}         &\colhead{LRD}}
\startdata
$a_{\rm x}\sin i$ (lt-s)&\dotfill     &53.4876                \\
Eccentricity            &\dotfill               &$<0.00004$             \\
$\dot{P}_{\rm pulse}/P_{\rm pulse}$ (yr$^{-1}$) &$-4.52(2)\times 10^{-4}$~\tablenotemark{a}
&$-5.30(2)\times 10^{-4}$\\
$a_{0}$\tablenotemark{b} ~(MJED)&42836.18278(20)        &42836.18277(20)        \\
$a_{1}$\tablenotemark{b} ~(days)&3.89229090(43) &3.89229118(48)                 \\
$a_{2}$\tablenotemark{b} ~(days)&$-6.953(28)\times 10^{-8}$  &$-6.97(3)\times 10^{-8}$\\
$\dot{P}_{\rm orb}/P_{\rm orb}$ (yr$^{-1})$ &$-3.353(14)\times
10^{-6}$~\tablenotemark{a}&$-3.36(3)\times 10^{-6}$\\
\enddata
\tablenotetext{a}{Evaluated at MJED 50000.}
\tablenotetext{b}{Center time of the $N^{\rm th}$ eclipse is given by $t_{\rm N}=a_{0}+a_{1}N+a_{2}N^{2}$.} 
\label{timing_results}
\end{deluxetable}
\clearpage

\begin{deluxetable}{lll}
\tablewidth{3.0in}
\tablecaption{Coefficients of the Quadratic Fit to the Pulse Period Data}
\tablehead{
\colhead{Parameter}     &\colhead{Value}&\colhead{} }
\startdata
$p_{0}$ &$0.707753$                     & s             \\
$p_{1}$ &$-8.75\times 10^{-6}$          &s day$^{-1}$   \\
$p_{2}$ &$1.68\times 10^{-11}$          &s day$^{-2}$   \\
$t_{0}$ &2450000.5                      &JD            \\
\enddata
\tablenotetext{}{$P_{\rm pulse}=p_{0}+p_{1}(t-t_{0})+p_{2}(t-t_{0})^{2}$.} 
\label{per_fit}
\end{deluxetable}
\clearpage

\figcaption[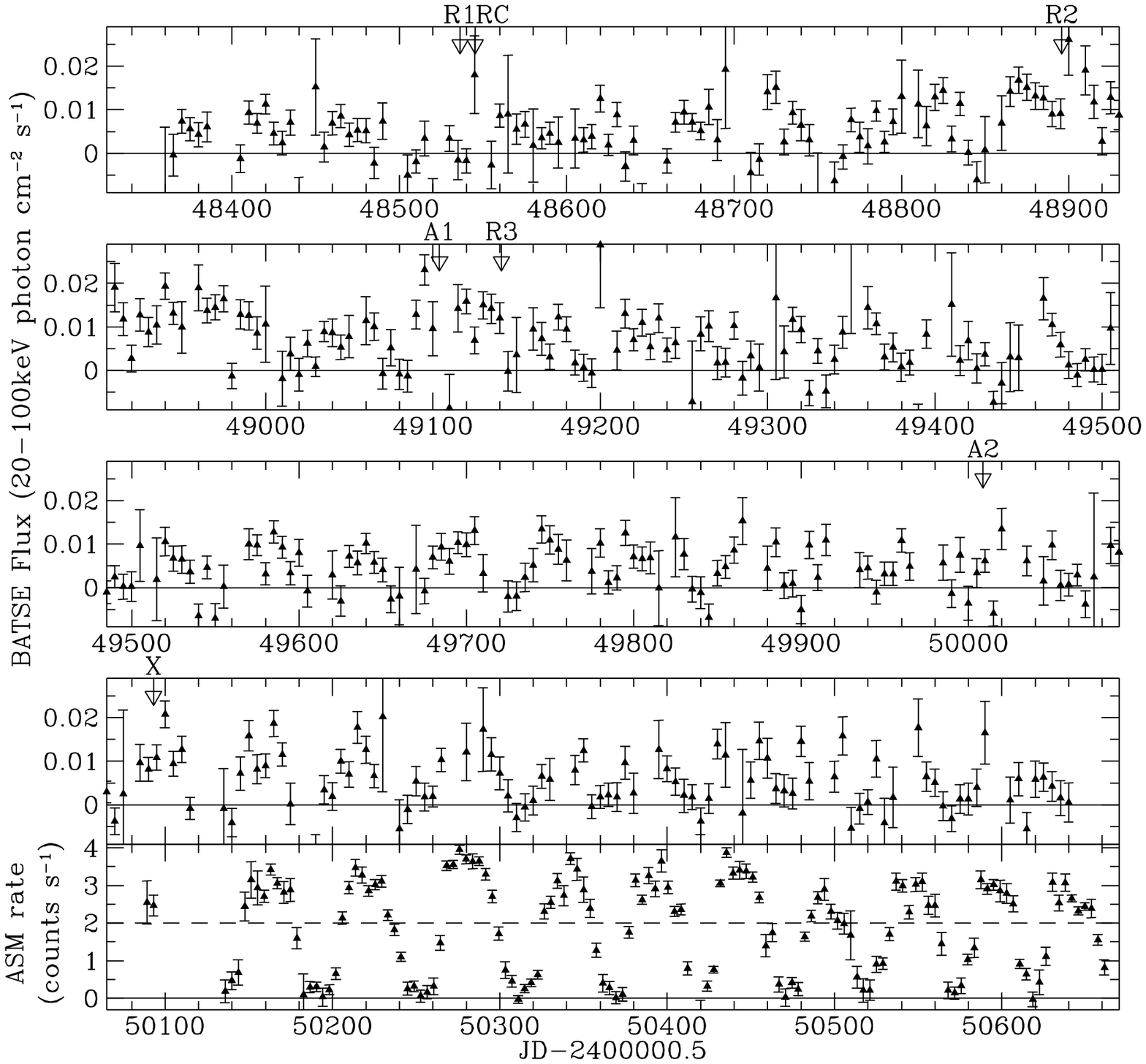]{{\it CGRO\/} BATSE
  and {\it RXTE\/} ASM (1.5-12 keV)light curves of SMC~X-1.  The time
  of the start of each of the pointed observations is indicated on
  this figure by R for {\it ROSAT\/}, A for {\it ASCA\/}, and X for
  {\it RXTE\/}.} 

\figcaption[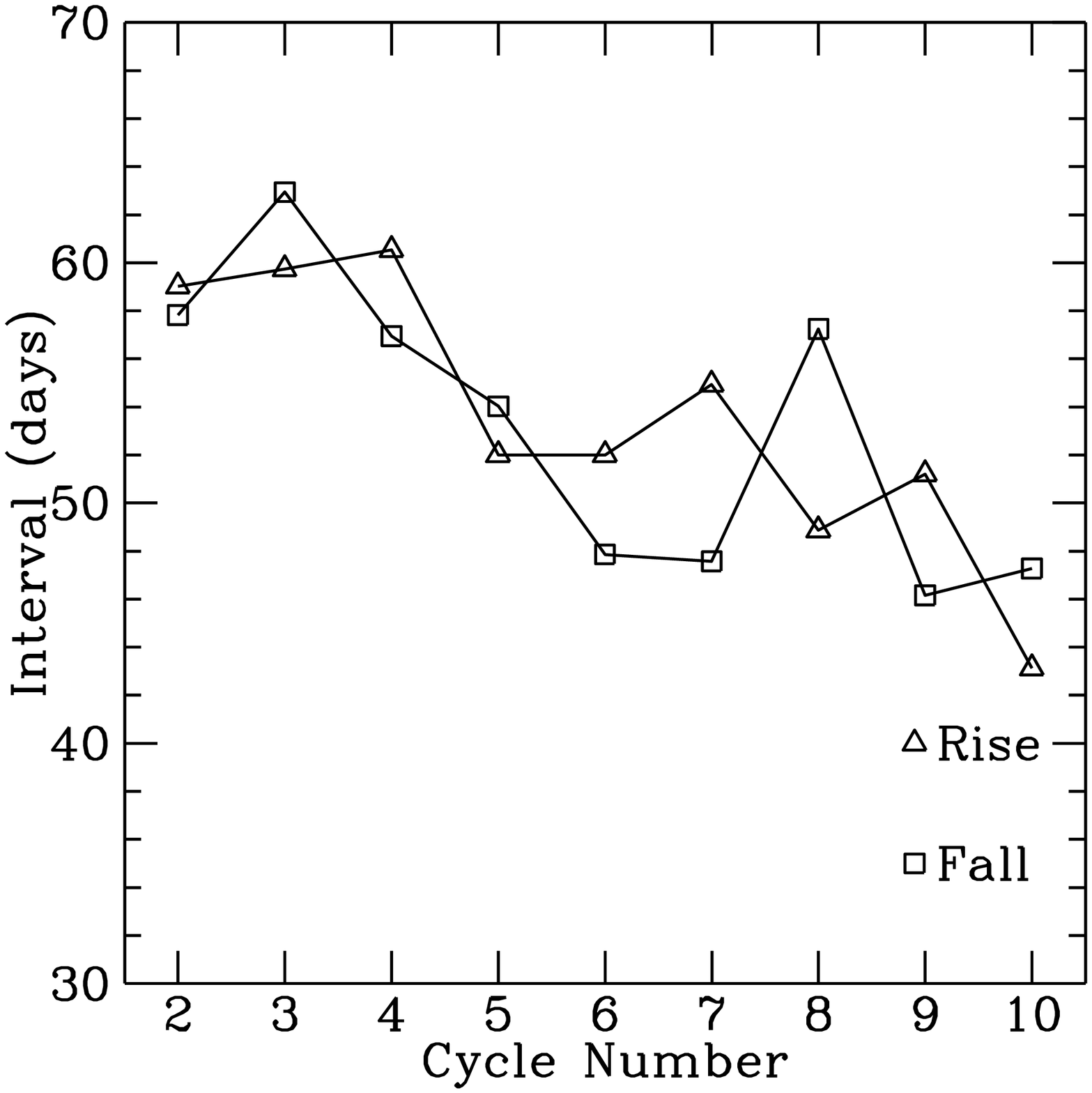]{Time intervals
  between successive high/low transitions of SMC~X-1.  The transition
  times are obtained from crossings of 2 counts s$^{-1}$ in the ASM
  light curve. 
  }  

\figcaption[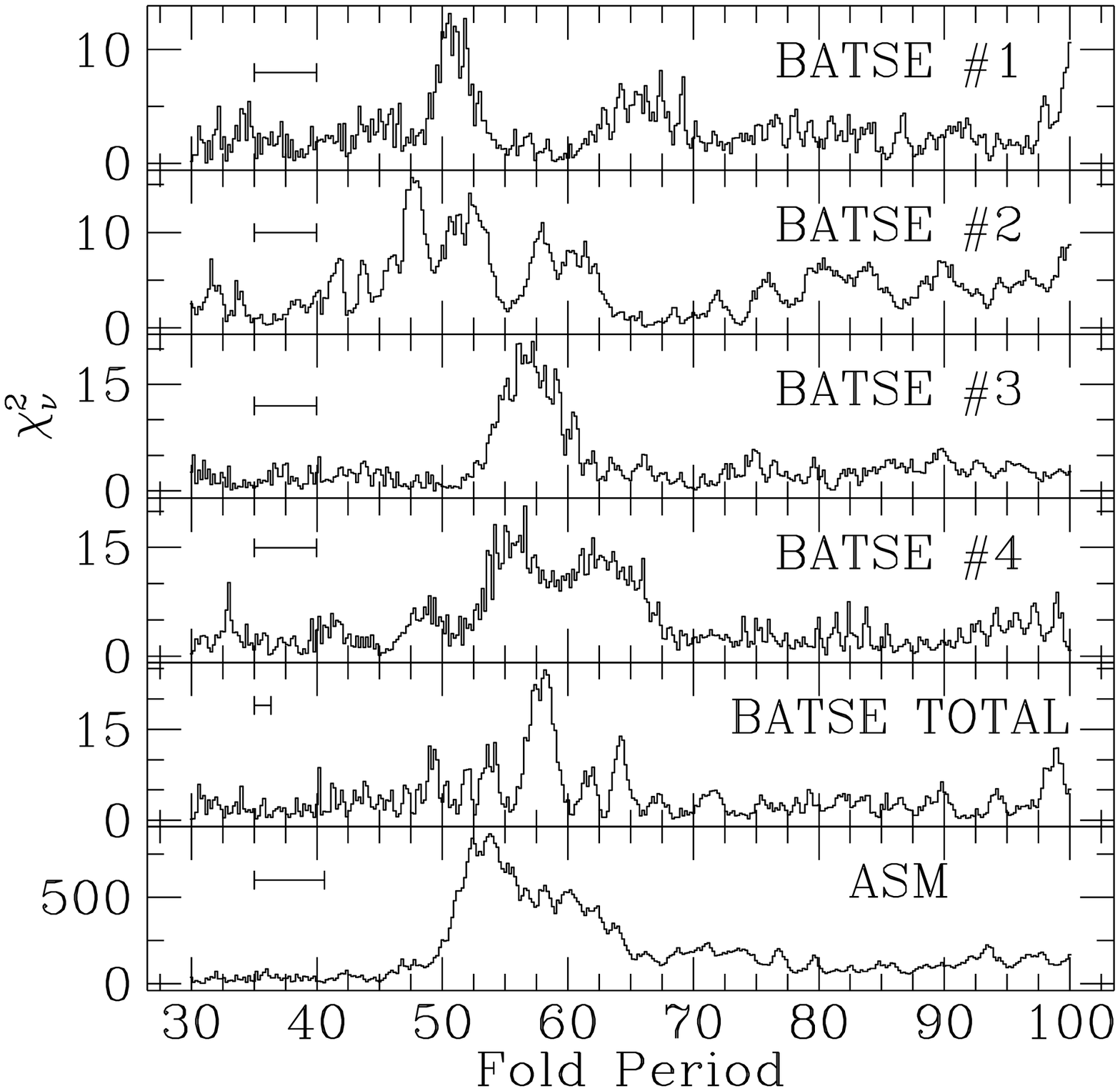]{Deviations
  from the average count rates, given in terms of a reduced
  chi-squared statistic, of four sections of the BATSE light
  curve, of the entire BATSE light curve, and the ASM light curve when
  folded at various trial periods.  The horizontal bars indicate the
  expected width (FWHM) of a peak corresponding to a sinusoidal
  variation with constant period.}  

\figcaption[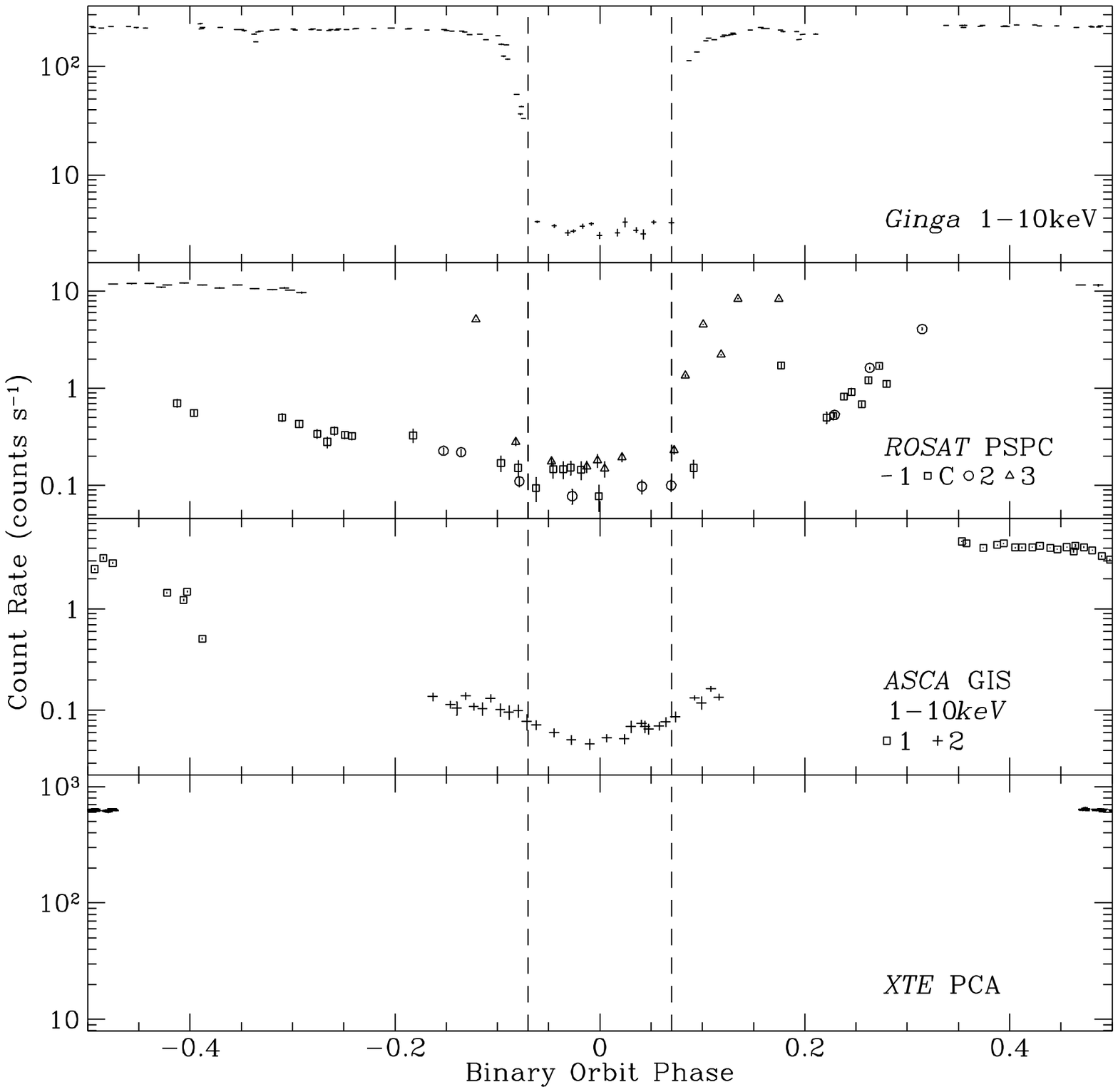]{Count rates of
  SMC~X-1 as a function of orbital phase for the pointed observations.
  The {\it ROSAT\/} PSPC region C count rates are corrected for off-axis
  vignetting.  Dashed lines at orbital phase $\pm$0.07 indicate the
  end of immersion and beginning of emersion of the eclipse
  transitions observed in the high state.} 

\figcaption[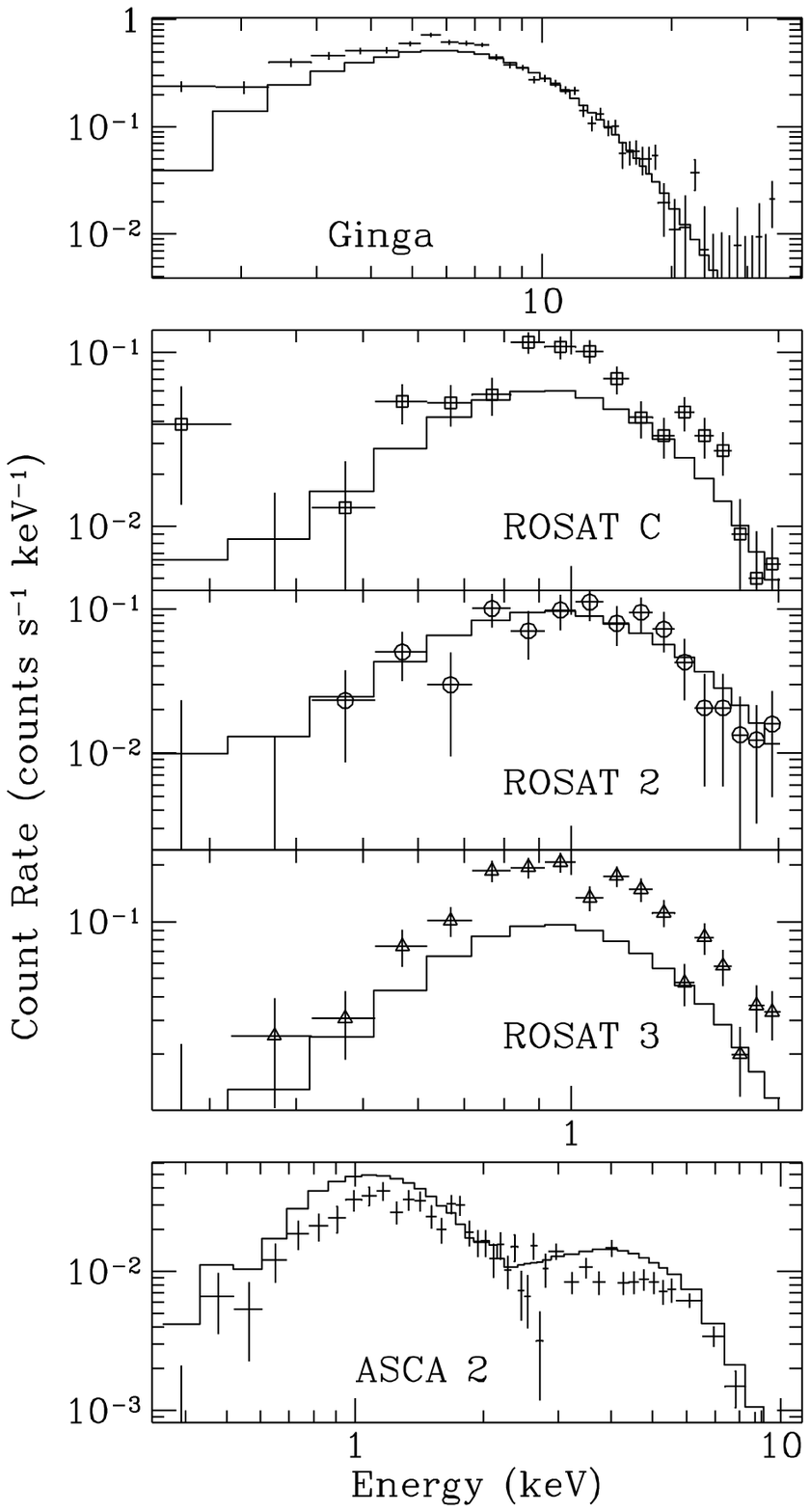]{Observed photon number spectra during eclipses
  (orbital phase in the range -0.07 to +0.07) and the computed spectra
(solid lines) for a single fitted model.}

\figcaption[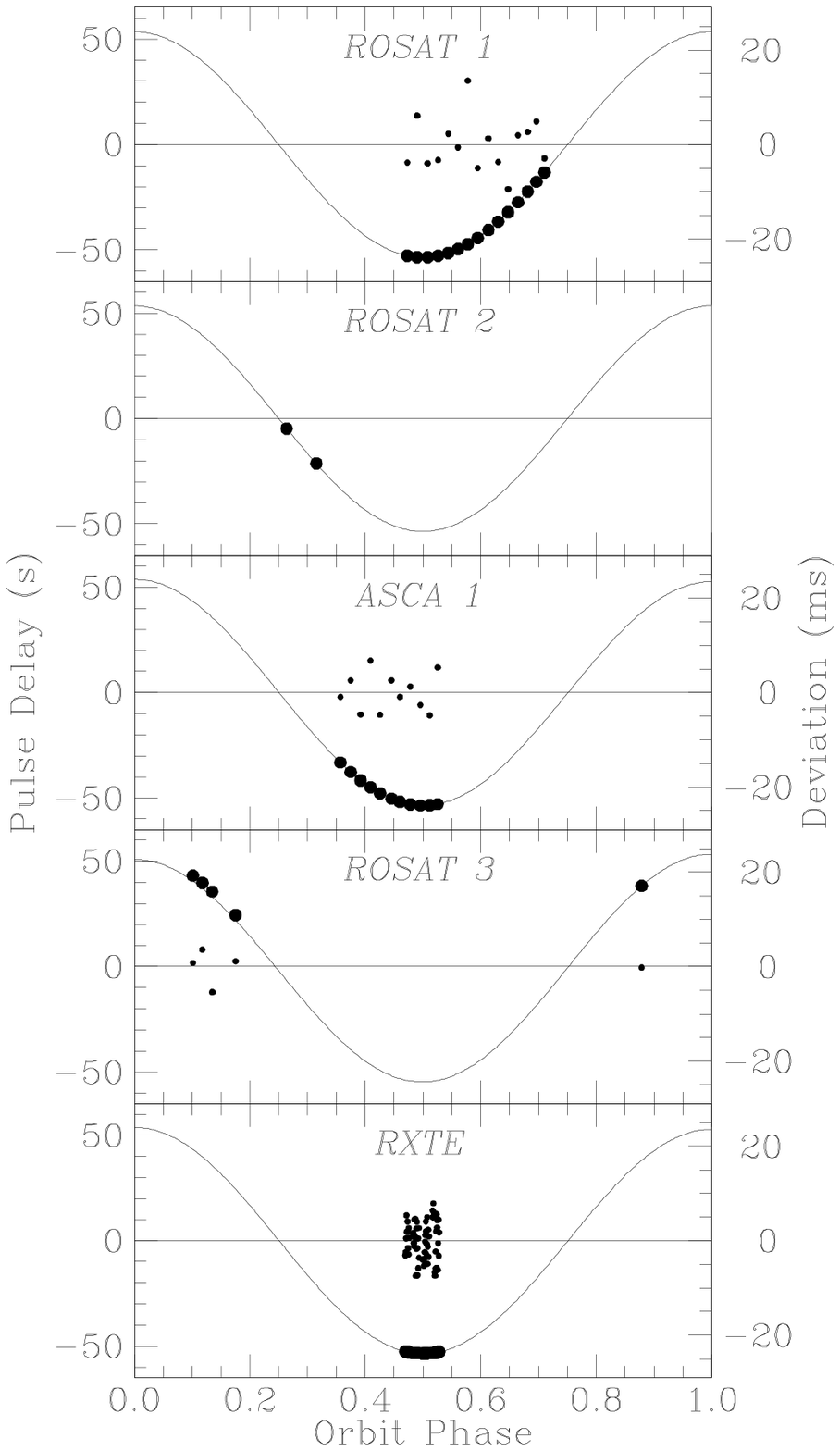]{Pulse arrival-time delays
  plotted against orbital phase together with the fitted curves of
  predicted delays and residuals.} 

\figcaption[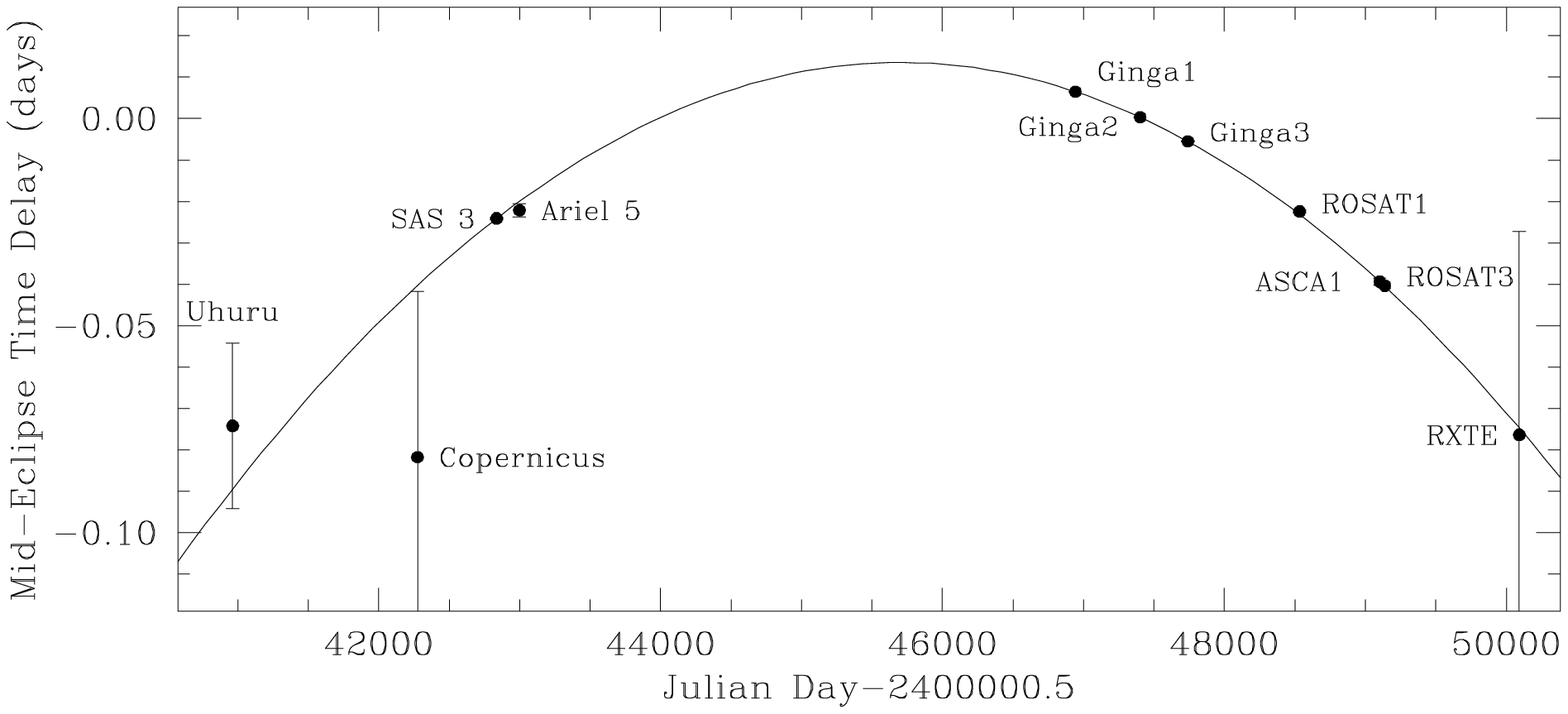]{Differences between
  the measured values of eclipse center times ($T_{\pi /2}$) and
  values computed for a constant orbital period ($P_{\rm
  orb}$=3.892188).  The solid curve is the fitted function (see Table
  5) which includes a linear decrease of the orbital period.}

\figcaption[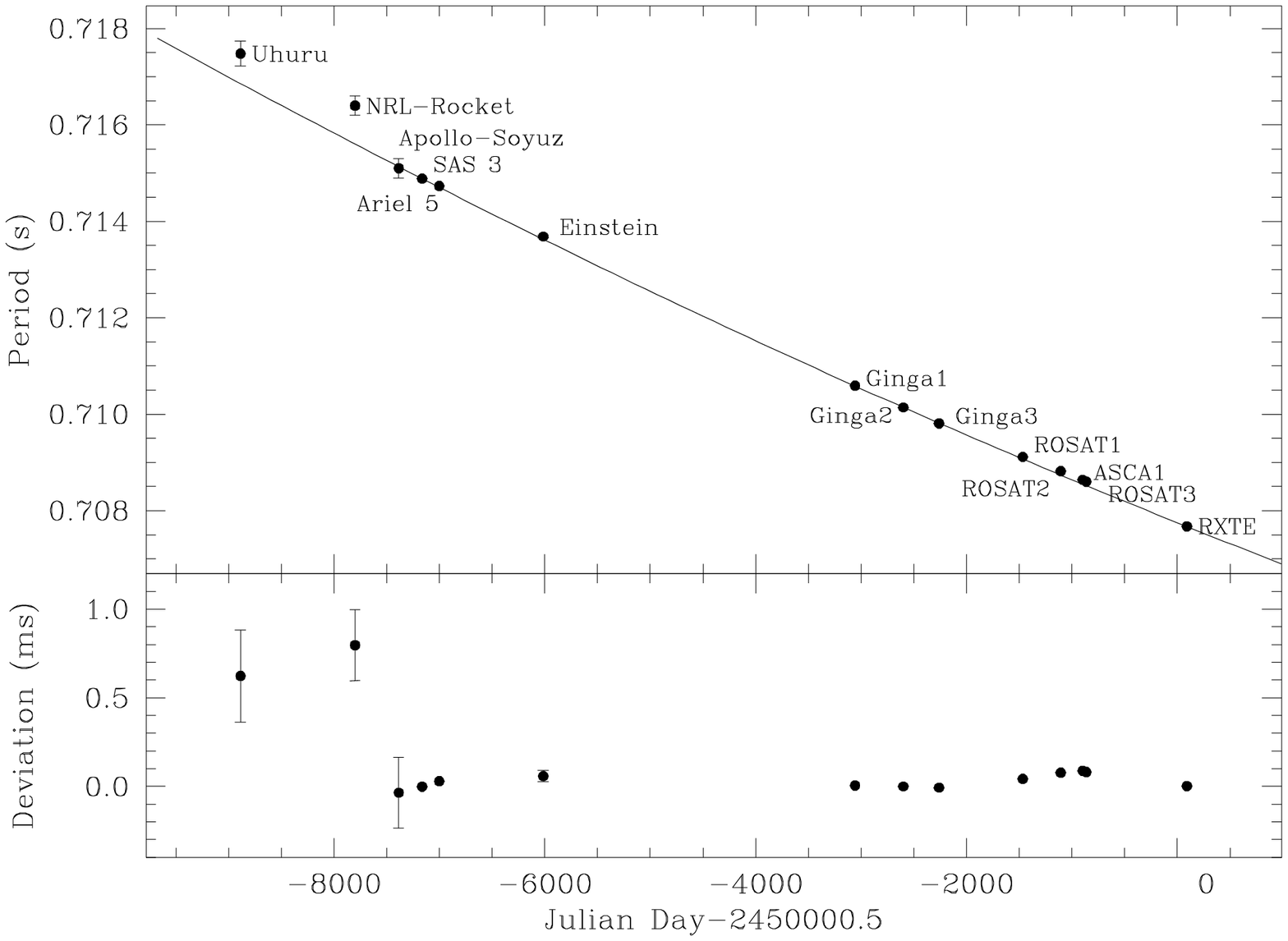]{Inferred intrinsic pulse
  period plotted 
  against time together with a fitted quadratic and residuals.}

\figcaption[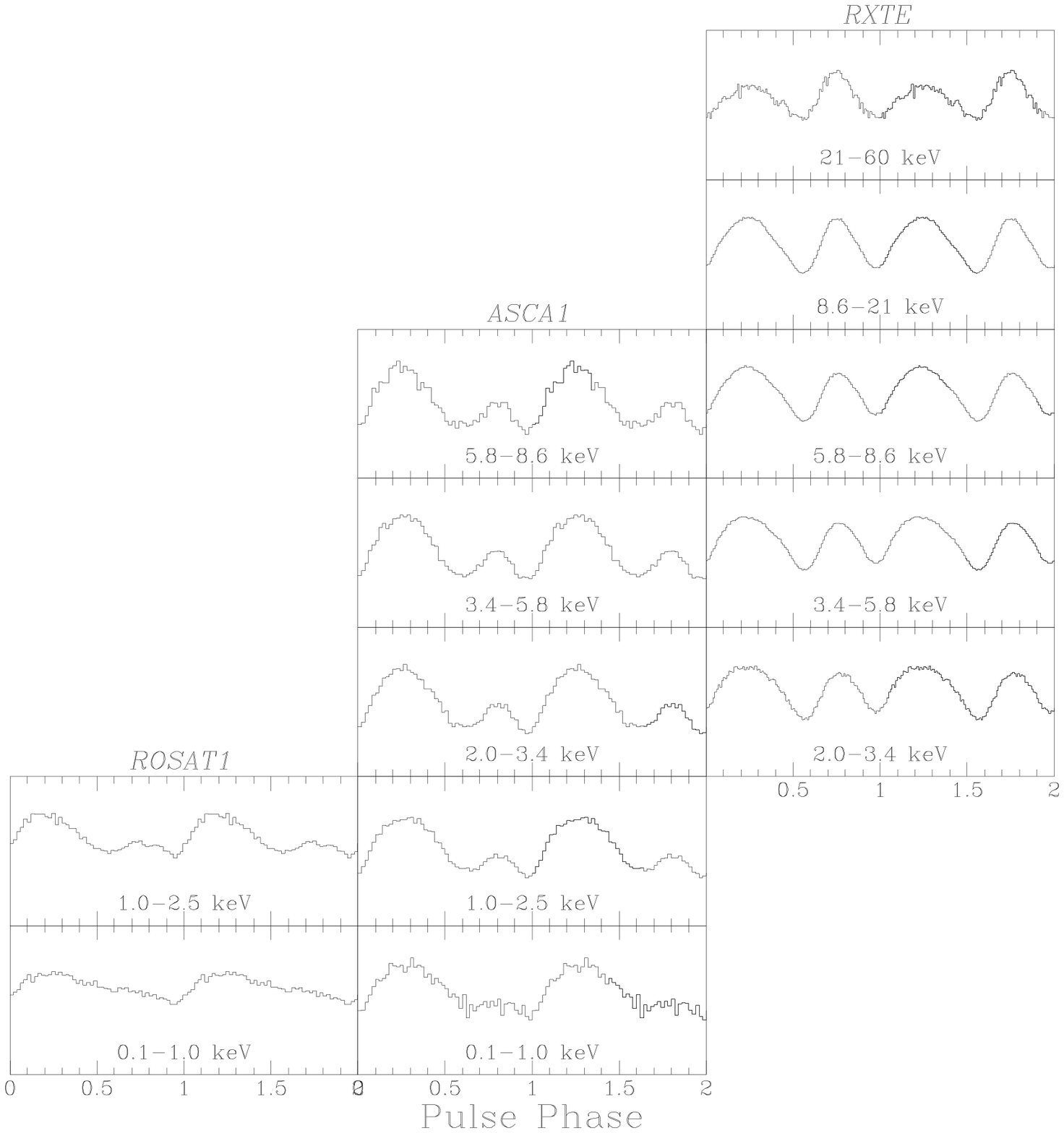]{Pulse profiles in various energy 
  bands derived from the {\it ROSAT} 1, {\it ASCA} 1, and {\it RXTE}
  data.}

\figcaption[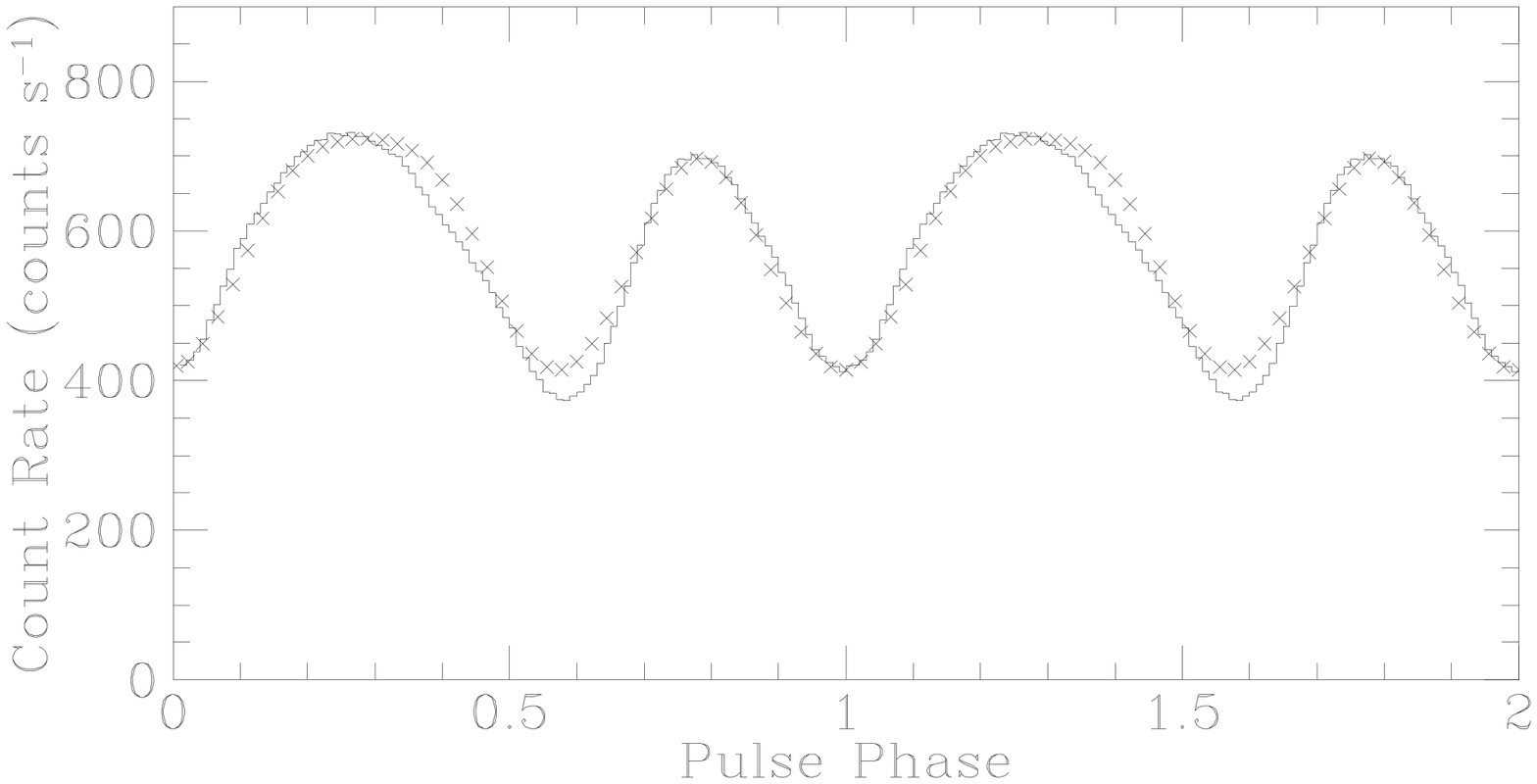]{Computed pulse profile
  (crosses) modeled by pencil and fan beams and fitted to the observed
  average 2-21 keV pulse profile (histogram) derived from the {\it
  RXTE\/} PCA data.} 

\figcaption[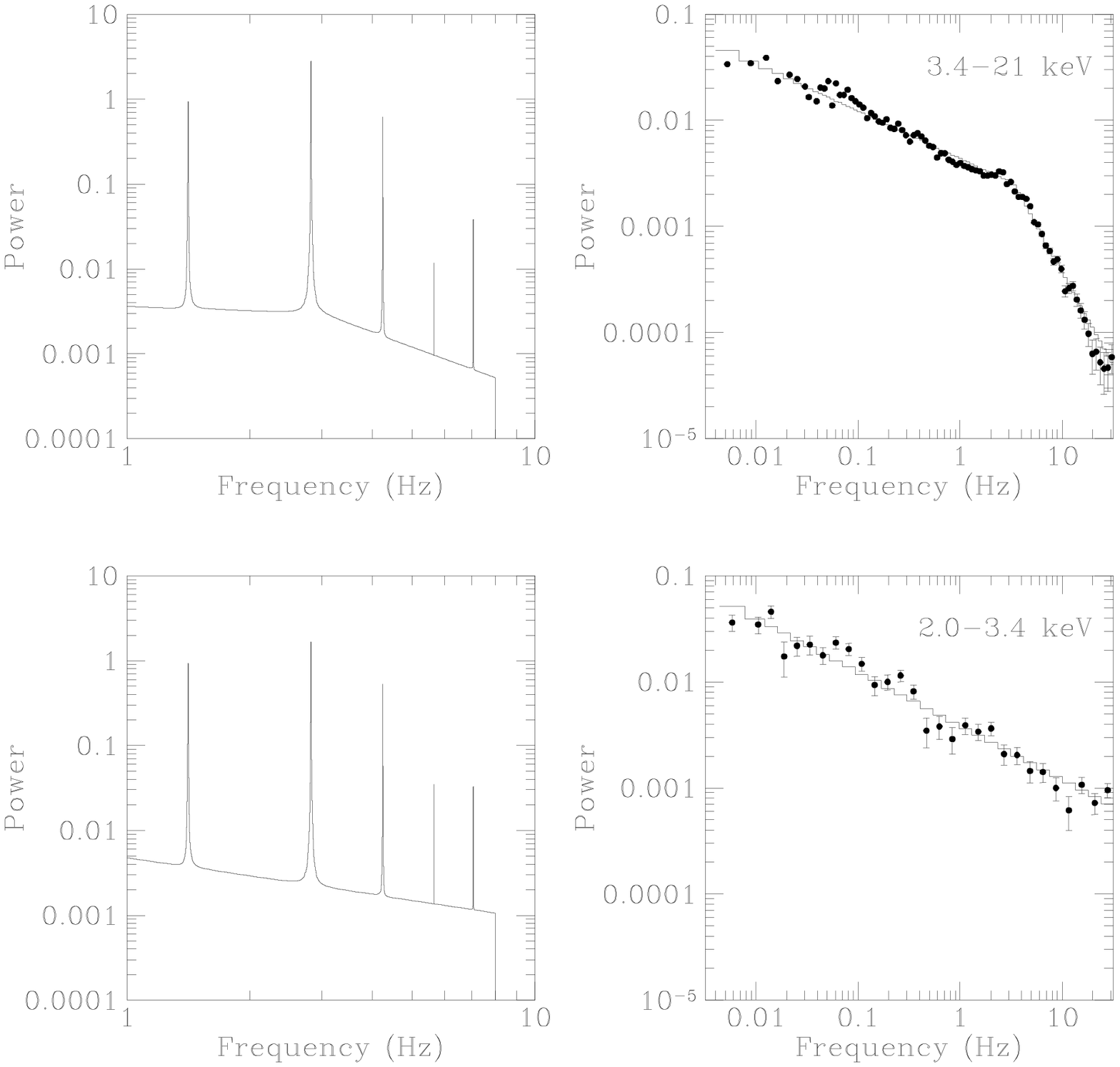]{Left panels: Fitted harmonic and
  continuum components of the power spectra in two energy bands of
  {\it RXTE\/} PCA data with times corrected for orbital motion. Right
  panels: Power spectra after subtraction of the harmonic components.} 

\clearpage
\plotone{fig01_BATSE_and_RXTE_monitor_data.ps}
\clearpage
\plotone{fig02_rise_fall_time_intervals.ps}
\clearpage
\plotone{fig03_rchisq_vs_fold_period.ps}
\clearpage
\plotone{fig04_ctrates_vs_orbit_phase.ps}
\clearpage
\plotone{fig05_spectra.ps}
\clearpage
\plotone{fig06_five_orbit_mrqfits.ps}
\clearpage
\plotone{fig07_eclipse_centers_vs_JD.ps}
\clearpage
\plotone{fig08_pulse_period_vs_time.ps}
\clearpage
\plotone{fig09_pulse_profiles.ps}
\clearpage
\plotone{fig10_pulsar_simulation.ps}
\clearpage
\plotone{fig11_xte_pwrspec.ps}

\end{document}